\UseRawInputEncoding
\documentclass[twocolumn, tighten, dvipsnames, floatfix]{aastex63}

\usepackage[]{xcolor}
\usepackage{booktabs}
\usepackage{physics}
\usepackage{wasysym}

\usepackage{longtable}
\usepackage{threeparttablex} 

\usepackage{array}
\usepackage{amsmath}
\usepackage{changepage}
\usepackage{lineno, ulem}
\usepackage[flushmargin]{footmisc}
\usepackage{afterpage}
\usepackage{subfigure}

\newcommand{\kms}{\mbox{$\>{\rm km\, s^{-1}}$}}

\usepackage{lipsum}
\usepackage{float}

\shorttitle{The Curious Case of BD+20 2457}
\shortauthors{Perottoni et al.}

\begin{document}

\title{Searching for Extragalactic Exoplanetary Systems: the Curious Case of BD+20 2457}

\correspondingauthor{H\'elio D. Perottoni}
\email{hperottoni@gmail.com}

\author[0000-0002-0537-4146]{H\'elio D. Perottoni}
\affil{Universidade de S\~ao Paulo, Instituto de Astronomia, Geof\'isica e Ci\^encias Atmosf\'ericas, Departamento de Astronomia, \\ SP 05508-090, S\~ao Paulo, Brazil}

\author[0000-0002-7662-5475]{Jo\~ao A. S. Amarante}
\altaffiliation{Visiting Fellow}
\affil{Jeremiah Horrocks Institute, University of Central Lancashire, Preston PR1 2HE, UK
}

\author[0000-0002-9269-8287]{Guilherme Limberg}
\affil{Universidade de S\~ao Paulo, Instituto de Astronomia, Geof\'isica e Ci\^encias Atmosf\'ericas, Departamento de Astronomia, \\ SP 05508-090, S\~ao Paulo, Brazil}

\author[0000-0002-5274-4955]{Helio J. Rocha-Pinto}
\affiliation{Universidade Federal do Rio de Janeiro, Observat\'orio do Valongo, Lad. Pedro Ant\^onio 43, 20080-090, Rio de Janeiro, Brazil}

\author[0000-0001-7479-5756]{Silvia Rossi}
\affil{Universidade de S\~ao Paulo, Instituto de Astronomia, Geof\'isica e Ci\^encias Atmosf\'ericas, Departamento de Astronomia, \\ SP 05508-090, S\~ao Paulo, Brazil}

\author[0000-0003-4524-9363]{Friedrich Anders}
\affil{Institut de Ci\`encies del Cosmos, Universitat de Barcelona (IEEC-UB), Carrer Mart\'i i Franqu\`es 1, 08028 Barcelona, Spain}

\author[0000-0003-3382-1051]{Lais Borbolato}
\affil{Universidade de S\~ao Paulo, Instituto de Astronomia, Geof\'isica e Ci\^encias Atmosf\'ericas, Departamento de Astronomia, \\ SP 05508-090, S\~ao Paulo, Brazil}

\begin{abstract}

Planets and their host stars carry a long-term memory of their origin in their chemical compositions.
Thus, identifying planets formed in different environments improves our understating of planetary formation. Although restricted to detecting exoplanets within the solar vicinity, we might be able to detect planetary systems that formed in small external galaxies and later merged with the Milky Way. In fact, \textit{Gaia} data have unequivocally shown that the Galaxy underwent several significant minor mergers during its first billion years of formation. The stellar debris of one of these mergers, Gaia-Enceladus (GE), is thought to have built up most of the stellar halo in the solar neighborhood.
In this Letter, we investigate the origin of known planet-host stars combining data from the NASA Exoplanet Archive with \textit{Gaia} EDR3 and large-scale spectroscopic surveys. We adopt a kinematic criterion and identify 42 stars associated with the Milky Way's thick disk and one halo star. The only halo star identified, BD+20 2457, known to harbor two exoplanets, moves on a retrograde and highly eccentric orbit. Its chemical abundance pattern situates the star at the border between the thick disk, the old halo, and accreted populations. Given its orbital parameters and chemical properties, we suggest that BD+20 2457 is likely formed in the protodisk of the Galaxy, but we do not exclude the possibility of the star belonging to the debris of GE. Finally, we estimate a minimum age and mass limit for the star, which has implications for its planetary system and will be tested with future Transiting Exoplanet Survey Satellite observations.

\end{abstract}



\section{Introduction}
\label{sec:intro}

\setcounter{footnote}{4}

The advent of surveys dedicated to searching for exoplanets covering large areas of the sky, e.g., Kepler mission \citep{borucki2010} and Transiting Exoplanet Survey Satellite (TESS) mission \citep{ricker2014}, has increased the number of exoplanet candidates to 4325 around 3216 host stars\footnote{According to NASA Exoplanet Archive in 2020 December.}. Consequently, 
the improvement of the statistical census of exoplanets (see \citealt{Adibekyan2019}) contributes to a better understanding of, e.g., ($i$) the dependence of planet properties (e.g. giant planet occurrence, orbital features, and planetary composition) with its host star's chemical abundance \citep[e.g.,][]{Gonzalez1997,fischervalenti2005,petigura2018,Teske2019}, ($ii$) the planetary formation in different environments (\citealt{Nayakshin2017,Brasser2020}), and ($iii$) the fraction of stars with exoplanets in our Galaxy (e.g., \citealt{ raymond2007,DressingCharbonneau2013,DressingCharbonneau2015, hsu2019}).

The universality of the aforementioned features are yet to be tested; it will be achieved only with the study of exoplanets in external galaxies. Up to now, only a few planetary-mass objects have been found in extragalactic systems through microlensing events (see \citealt{dai2018,bhatiani2019,distefano2020}). This is mainly due to the current impossibility to use the traditional detection methods, radial velocity, and transit, in resolved stellar sources at large distances. In principle, it could be done in the future for, e.g., M31 \citep{ingrosso2009}, Large Magellanic Cloud (LMC) \citep{lund2015}, and Small Magellanic Cloud (SMC) \citep{mroz2017}. Nonetheless, we do not need to search for exoplanets in nearby galaxies to study extragalactic planetary systems: the Milky Way (MW) may already host exoplanets from accreted dwarf galaxies.

The Galaxy experienced a significant accretion event circa $\sim$10 Gyr ago (at redshift $z \sim 2$) with a dwarf galaxy named Gaia-Enceladus (GE; \citealt{helmi2018,belokurov2018}). The stars accreted from this event constitute most of the nearby/inner halo, i.e., within at least 3 kpc from the Sun (e.g., \citealt{Haywood2018,DiMatteo2019}). \citet{carrillo2020} used a sample of TESS candidates cross--matched with \textit{Gaia} and showed that $\sim$1$\%$ TESS target stars inhabit the Galactic stellar halo (hereafter ``halo"; their figure 4 clearly shows the kinematic signature of GE). This opens the possibility to investigate extragalactic exoplanets in stars that were accreted during this event.

With this motivation, we investigated the dynamical properties of known exoplanet-host stars looking for systems with halo-like orbits. With data from \textit{Gaia} Early Data Release 3 (GEDR3) and large-scale ground-based spectroscopic surveys, we found a strong candidate which we further investigate according to its abundances. The paper is organized as follows. Section \ref{sec:data} describes the data used in our analysis; Section \ref{sec:ana} shows the kinematics, orbital parameters, and abundances of the exoplanet--host stars that likely belong to the MW's thick disk and halo; Section \ref{sec:BD} describes the BD+20 2457 system and discuss its possible chemo-dynamical origin and implications to its planetary system; finally, in Section \ref{sec:conc}, we provide our conclusions and final remarks. We also include three appendices. Appendix \ref{app:qcc} describes the quality control applied to the data, Appendix \ref{app:formalism} describes the traditional kinematical selection criteria according to \citet{bensby2003}, and  Appendix \ref{app:exoplanets} presents a table of exoplanets classified as thick disk or halo and some of their features.

\section{Data} 
\label{sec:data}
This Letter is based on a combination of data from the Exoplanet Archive with APOGEE DR16, GEDR3, GALAH DR3, and LAMOST DR5 (\citealt{LAMOST1,exoplanetdatabase2013,APOGEE2020,gaiaedr3,galah2020} see Appendix \ref{app:qcc} for details about the sample selection criteria). We build four samples of planet-host stars containing the five astrometric parameters and their uncertainties from the GEDR3 catalog and radial velocities and errors from either the aforementioned spectroscopic surveys or \textit{Gaia} itself. Our combined sample contains 1559 planet-host stars and the individual samples from APOGEE, {\it Gaia}, GALAH, and LAMOST have 682, 919, 116, and 453 stars, respectively. In addition to that, we selected a cleaned sample of stars from APOGEE DR16 (see Appendix \ref{app:qcc}) to illustrate some of the features that will be presented. We refer to it as the \textit{APOGEE sample} throughout the Letter.\par
We calculated the orbits of the stars with the publicly available Python library \texttt{AGAMA} \citep{agama} for $\sim$5 Gyr forward. The Galactic potential model employed is described in \citet{mcmillan2017}. We adopt the values from \citet{Bovy2020} for the solar Galactocentric distance $R_{\odot}=8.22$, the local circular velocity $v_c = 243.0$\kms, and the solar motion with respect to the local standard of rest $(U_{\odot}, V_{\odot}, W_{\odot}) = (11.10,7.20,7.25)$\kms. We tested with other sets of fundamental Galactic parameters, e.g. those recommended by \citet{mcmillan2017}, and concluded that our main results do not depend on these choices.

For each star, we performed 1000 Monte Carlo realizations of the orbit according to Gaussian distributions of its uncertainties in the phase-space coordinates. The medians of each kinematic/dynamical quantity considered are taken as our nominal values. The adopted uncertainties are the 16th and 84th percentiles of the resulting distributions. In this work, our analyses are based on the complete velocity vector ($v_R$, $v_{\phi}$, and $v_z$) of each star in the cylindrical coordinate system (radial, azimuthal, and vertical directions, respectively), the maximum distance from the Galactic plane achieved during a star's orbit ($Z_{\max}$), the perigalactic ($r_{\min}$) and apogalactic ($r_{\max}$) distances, eccentricity ($e=(r_{\max}-r_{\min})/(r_{\max}+r_{\min})$), total orbital energy ($E$), and vertical component of the angular momentum $L_z = R_{\rm Gal} \times v_{\phi}$, where $R_{\rm Gal}$ is the plane-projected distance of a given star from the Galactic center.
\begin{figure}
\includegraphics[width=\columnwidth]{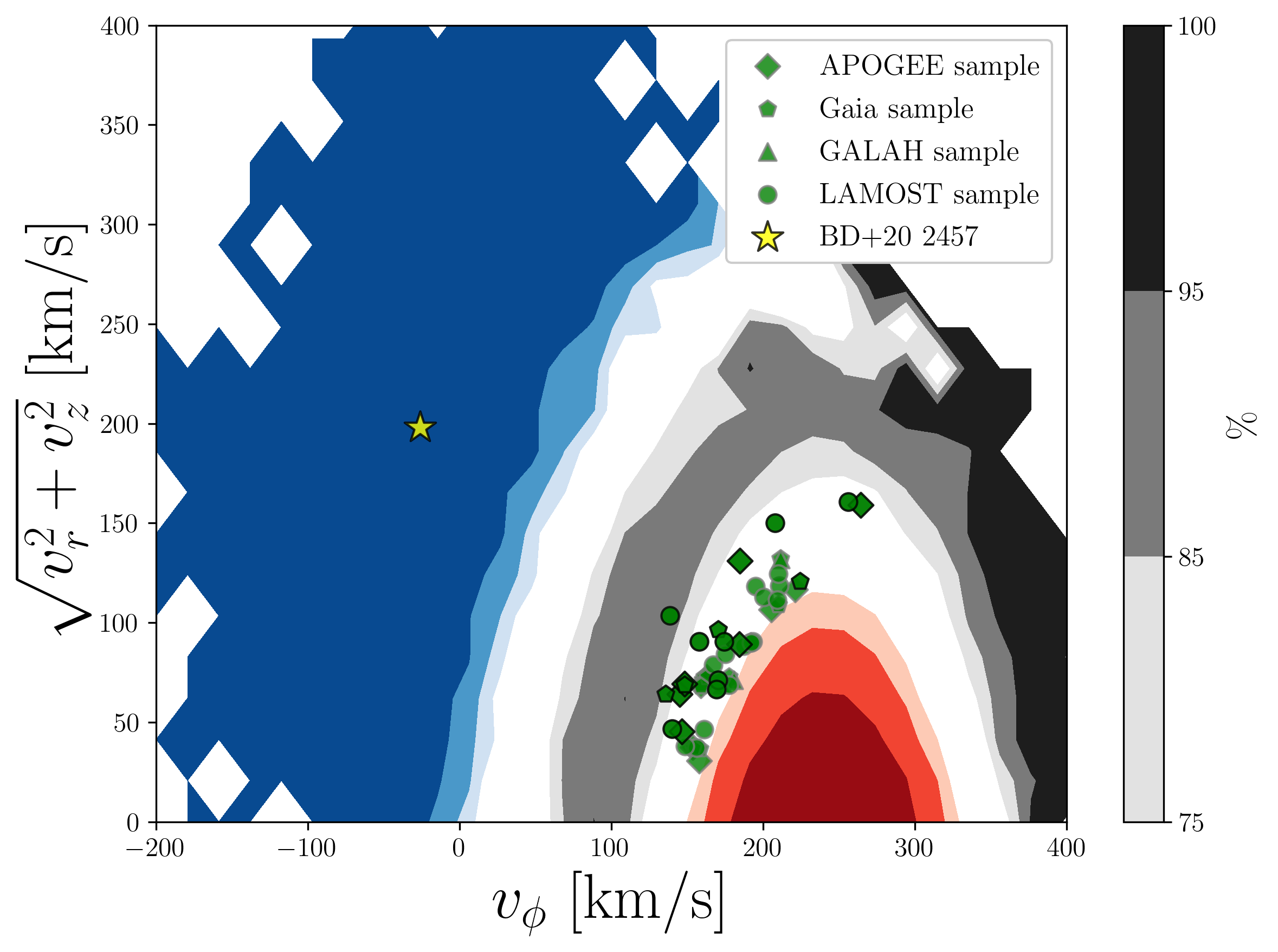}
\caption{Toomre diagram of the halo/thick-disk host stars. The thin disk, thick disk, and halo are represented by the red, gray and blue colours, respectively. The darkness in
shading corresponds to the fraction of a given population at that location of the diagram (75\%, 85\% or 95\%).
The host stars classified as thick-disk and halo stars are shown in green and yellow, respectively.  
The symbols with a black contour are the host stars classified as thick-disk stars ($TD/D > 10$) according to the kinematic criteria of \citet{bensby2003}.}
\label{fig:toomre}
\end{figure}
\section{Analysis}\label{sec:ana}

\begin{figure*}[ht]
\includegraphics[width=18.0cm]{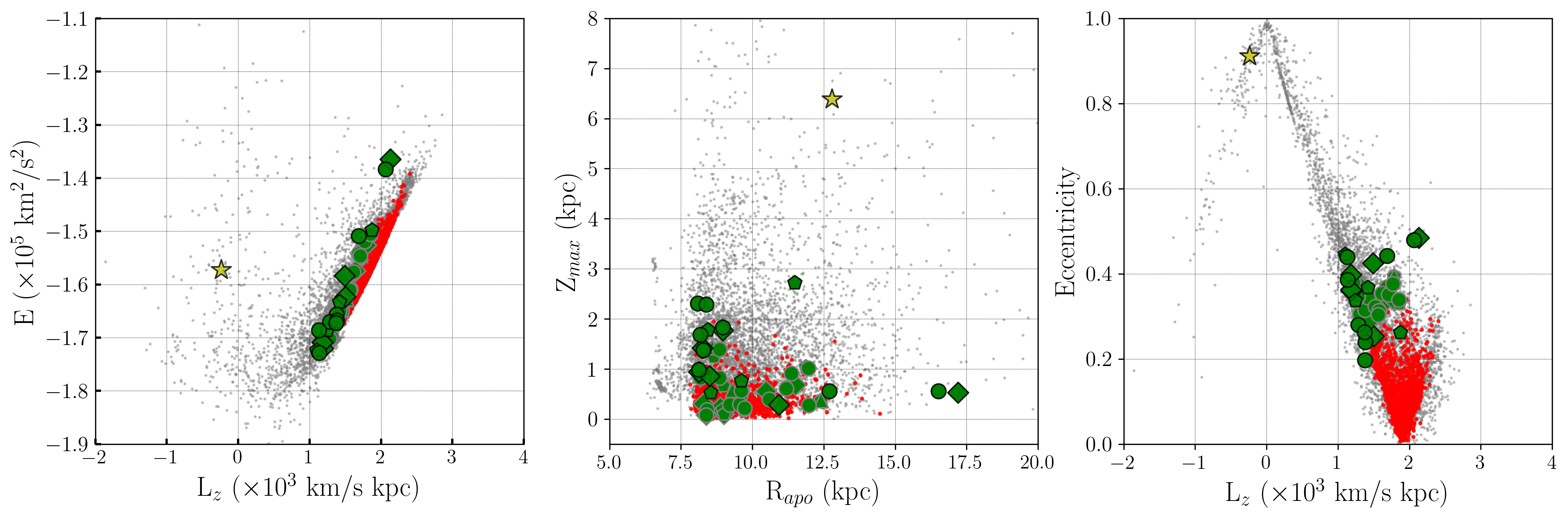}
\caption{The orbital parameters of host stars in the E--$L_Z$ (left), $Z_{\max}$--$r_{\max}$ (middle), and eccentricity--$L_{z}$ (right) planes.
The halo host star BD+20 2457 is represented as a yellow star symbol.
The host stars classified as belonging to the thick and thin disk are shown in green and red symbols, respectively.
The gray dots are stars from the \textit{APOGEE sample}, within a heliocentric radius of 2 kpc, which we used to illustrate the local stellar distribution. The green symbols are defined as in Figure \ref{fig:toomre}}.
\label{fig:orbitparam}
\end{figure*}
The Toomre diagram is generally used to kinematically classify halo and/or thick-disk stars (e.g., \citealt{bensby2003}). Given assumptions on the properties of the stellar populations (namely, their relative fractions and velocity distributions), we can predict which regions of the $v_{\phi}$ vs. $\sqrt{v_R^2 + v_z^2}$ plane will be dominated by a given population. We used the Galaxia model \citep{sharma2011} --- with an updated description of the thick disk and stellar halo velocity distributions and their local fractions (as described in \citealt{amarante2020b}) --- to define the regions of the diagram where each population dominates. This is illustrated in Figure \ref{fig:toomre}. The thin disk, thick disk, and halo are represented by the red, grey, and blue shaded areas, respectively. The darkness in the shading corresponds to regions where a given population has a fraction greater than 75\%, 85\% or 95\% overall. According to our criteria, thick-disk stars are all the objects that lie outside of the thin-disk predominant region. We used this diagram to initially select stars in our sample that are likely to belong to the halo.\par
Overplotted in Figure \ref{fig:toomre}, we show only the stars in our sample that were not classified as thin disk. According to our criteria, we identified 42 planet-host stars as thick-disk candidate stars\footnote{Stars classified as thick disk in more than one sample are counted as one object.} (see Table in Appendix \ref{app:exoplanets}). 
We also applied a formalism similar to \citet{bensby2003} (see Appendix \ref{app:formalism}) to classify our thick-disk candidates according to a second kinematic criterion. The stars classified as thick disk according to this method, those having thick-disk-to-thin-disk (TD/D) membership ratios higher than 10 
, are shown with a black contour in Figure \ref{fig:toomre}. Our final list contains 18 highly probable thick disk and 1 halo candidates. Thick-disk planet-host stars were previously confirmed to exist (e.g., \citealt{gan2020} for a recent discussion) and are important constraints to the planetary formation theory, but in this Letter we focus on the host star having halo-like characteristics.\par
The only star we found in the halo-dominated region results in a fraction of 0.06\% of halo planetary host stars in our sample. While this relatively fraction seems low, we note that TESS is expected observe 0.16\% of halo stars \citep{carrillo2020}. If halo stars have the same probability of disk stars to host a planet, $\sim$30\% (e.g., \citealt{zhu2021}), finding only one exoplanetary system in the halo is within the expectations given our sample size.\par

Figure \ref{fig:orbitparam} shows the orbital properties of the host stars. We also included as gray dots the \textit{APOGEE sample}, within 2 kpc from the Sun, for comparison. The left panel shows the $E$--$L_Z$ plane, where the thick-disk host stars (green circles) have smaller $L_z$ compared to the thin-disk ones (red circles), at the same $E$. The middle panel shows that, as expected, thick-disk host stars have larger vertical excursions from the plane ($\langle Z_{\max} \rangle =  1.28$ kpc) compared to thin-disk ones ($\langle Z_{\max} \rangle = 0.32$ kpc). Finally, the eccentricity--$L_{z}$ plane (right panel) shows that the majority of thick-disk host stars have lower average vertical component of the angular momentum and higher eccentricity ($\langle L_z \rangle = $ 1.43 $\times 10
^{3}$ \,kpc\,km\,s$^{-1}$; $\langle e \rangle = 0.35$) compared to the thin disk ($\langle L_z \rangle =$  1.87 $\times 10^{3}$ \,kpc\,km\,s$^{-1}$; $\langle e \rangle = 0.11$). All the aforementioned properties are in line with expectations given the kinematic characteristics of both populations and in agreement with previous studies that investigated the chemically defined thin- and thick-disk properties (see \citealt{LiZhao2018}).

Interestingly, the host star BD+20 2457 has kinematics and dynamics typical of a halo star (yellow star symbol in Figures \ref{fig:toomre} and \ref{fig:orbitparam}). It is on a retrograde orbit, $L_z = -235$\,kpc\,km\,s$^{-1}$, with high eccentricity, $e = 0.91$, and it reaches $Z_{\max} = 6.39$\,kpc from the Galactic plane. Besides, BD+20 2457 is located in a region of $E$--$L_Z$ space dominated by accreted stars (see \citealt{Helmi2020} for a review). The recently discovered GE is the most prominent kinematic stellar structure that overlaps with this star in all the orbital spaces (see \citealt{limberg2021} and references therein), including a stricter criterion: \cite{Feuillet2020} recommend a cut on the radial action of stars,\footnote{The radial action $J_R$ represents the radial excursion of a given star in an axisymmetric potential \citep{BinneyBook}.} $\sqrt{J_R}> 30$\,kpc\,km\,s$^{-1}$, to select a ``pure" GE sample. We find a radial action of $\sqrt{J_R}= (29.9\pm0.2)$\,kpc\,km\,s$^{-1}$, which means that BD+20 2457 is compatible with such a conservative criterion at the 1$\sigma$ level.

\section{BD+20 2457: An Intriguing Star}\label{sec:BD}

Having found the most likely host-star candidate to be originated from an accreted galaxy, BD+20 2457, we now proceed to explore its elemental abundances and discuss its origin and implications toward this particular star-planet system. 

\subsection{Chemistry}\label{sec:BDchem}

\begin{figure}
\includegraphics[width=\columnwidth]{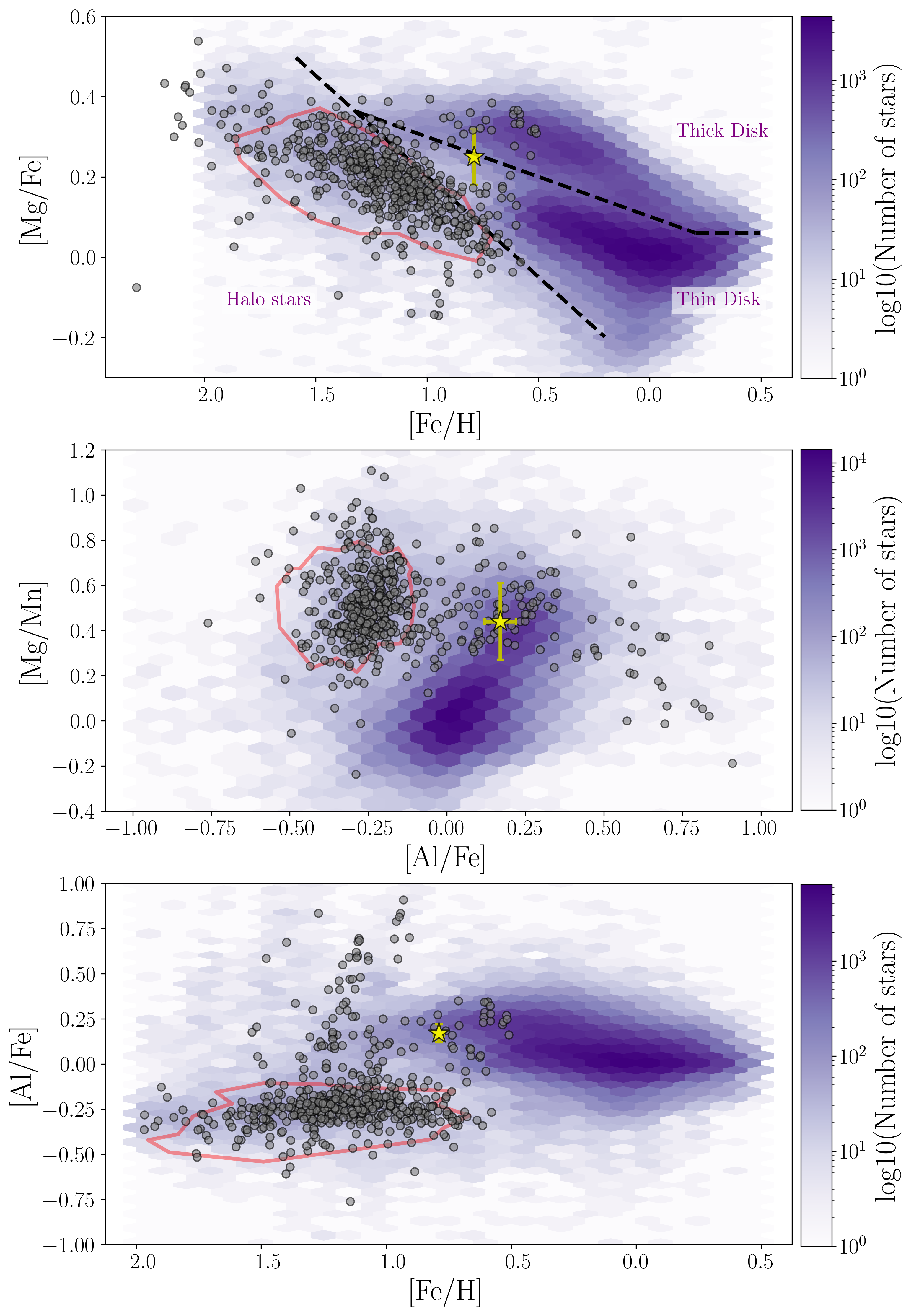}
\caption{The chemical abundances of the host star BD+20 2457 (yellow star) in the [Fe/H] and [Mg/Fe] (top), [Mg/Mn] and [Al/Fe] (middle), and [Al/Fe] and
[Fe/H] (bottom). The gray dots are GE candidate stars selected according to \cite{Feuillet2020} from the \textit{APOGEE sample} cut in $\rm[Fe/H] < -0.5$.
The black dashed line is used to separate three main stellar groups according to \cite{mack2019}.  The red contour indicates the ``blob" of accreted stars \citep{das2020}. The 2D histogram color coded background shows the distribution of the \textit{APOGEE sample} stars. 
}
\label{fig:abundances}
\end{figure}

We explore the elemental abundances of BD+20 2457 to investigate whether its chemical profile is also similar to that from GE. 
We adopted the [Fe/H], [Mg/Fe], [Mg/Mn], and [Al/Fe] from the high-resolution (
$R \geq 67000$) spectroscopic study of \cite{Maldonado2013}.
Figure \ref{fig:abundances} shows the abundance planes which are commonly used to depict accreted stars from \textit{in situ} populations. The dashed lines in the top panel segregate stars into thin disk, thick disk, and accreted halo following \citet{mack2019}. The red contours in all the panels define the chemistry associated with accreted stars according to \citet{das2020}. For comparison, we also show the \textit{APOGEE sample} as the 2D histogram. Despite having halo-like kinematics and dynamics, BD+20 2457 has a chemical pattern consistent with the regions dominated by the chemical thick disk.\par

However, we cannot immediately rule out the possibility that BD+20 2457 was born in the GE's progenitor system since this substructure is expected to have a large spread in abundances \citep{matsuno2019}, commensurate with a massive progenitor. In fact, in Figure \ref{fig:abundances}, we also show that a small number of GE stars (gray dots) that were selected from the \textit{APOGEE sample} ($\rm[Fe/H] < -0.5$), following \cite{Feuillet2020}'s conservative selection criteria, have chemical properties similar to BD+20 2457 and thick-disk stars.\par
With the ambiguity brought from the chemodynamical analysis of BD+20 2457 star, we proceed with a more detailed investigation on the origins of this star.
\subsection{Accreted or Heated?}\label{sec:BDacc}
The accretion event of the massive dwarf galaxy GE (\citealt{belokurov2018,helmi2018}) occurred approximately $\sim$10 Gyr ago (\citealt{gallart2019, montaban2020}) and is thought to be responsible for the formation of the thick disk either by dynamical heating (\citealt{bignone2019,DiMatteo2019,gallart2019}) or by an induced star-formation burst (\citealt{bignone2019,grand2020}).

The stellar debris from this merger dominate the inner halo (\citealt{Haywood2018,DiMatteo2019,naidu2020}), although the latter also has a more metal-rich counterpart proposed to be formed \textit{in situ} (\citealt{bonaca2017,gallart2019}) and heated to halo-like orbits \citep{DiMatteo2019,gallart2019,amarante2020b,grand2020}.
\cite{belokurov2020} suggested that the metal-rich portion of the inner halo is a distinct component of the Galaxy and called it ``Splash".
According to this hypothesis, the Splash was part of the protodisk of the Galaxy that was dynamically heated to halo-like orbits. 

In this context, the metal-rich component of the inner halo has $v_{\phi}$, metallicity distributions, and several orbital parameters (\citealt{belokurov2020}; see their figure 1) that are in agreement with the values obtained for BD+20 2457. The main difference is that the nearby Splash stars typically do not reach values of  $Z_{\max} > 4$ kpc. Thus, the scenario of a merger event that dynamically heated the protodisk appears to explain the puzzling thick-disk-like chemical composition and the halo-like orbit of the BD+20 2457. Alternatively, \citet{amarante2020a} pointed out that a Splash-like population, i.e., a heated thick disk, could already be present in the young Galaxy during the GE merger event due to an early clumpy formation. Although it adds an extra puzzle regarding the BD+20 2457 origin, the clumpy formation would have occurred during the first Gyrs of the Galaxy \citep{clarke2019} and thus is also able to constrain the BD+20 2457's age as will be discussed in Section \ref{sec:arr}. \par
In Section \ref{sec:BDchem}, we showed that BD+20 2457 has abundances typical of the thick disk, but also that some dynamically selected GE candidates overlap with the thick disk in chemical parameter space. This is also seen in \cite{Matsuno2021} (see their figure 4) for the [Mg/Fe] ratio from a GE sample selected from GALAH DR3. In an attempt to quantify the rarity of stars with the same chemical and orbital properties of BD+20 2457, we search in the \textit{APOGEE sample} for stars that have [Fe/H], [Mg/Fe], [Mg/Mn], and [Al/Fe] within an interval of $\pm$0.1 dex from the abundances of BD+20 2457. Although not perfect, such monoabundance selection enables the study of coeval populations properties (e.g., \citealt{bovy2012, beraldo2020}). We found that 1270 stars satisfy these criteria but only 3 stars (0.2$\%$) were dynamically selected as GE candidates \citep{Feuillet2020}. This may indicate that protodisk stars when heated typically do not have GE-like orbits or that the fraction of GE contamination in the chemical abundance space is very small. In either case, further investigations are necessary to estimate the fraction of stars with these characteristics that could be related to either GE or Splash. \par
We also estimated how likely is BD+20 2457 to belong either to the thick disk or GE based on the ([Fe/H], [Mg/Fe], [Al/Fe], [Mg/Mn]) chemical space.  We calculated a multivariate KDE with the {\tt R}-package {\tt ks} \citep{RBook}, and derived a chemical thick-disk-to-GE likelihood ratio. This is a similar approach used by \cite{bensby2003} in the kinematic space. In our analysis, the thick-disk sample was selected according to \cite{Mackereth2019}, shown in Figure \ref{fig:abundances}, and the GE candidate stars were selected according to \cite{Feuillet2020}. We found that BD+20 2457 has a chemical thick-disk-to-GE likelihood ratio $\gg 1000$, which reinforces our preliminary analysis that it is chemically more akin to the thick disk. \par
Finally, we can exclude that any sort of bar interaction could be the cause of the retrograde motion and eccentric orbit observed for this star. \citet{fiteni2021} showed that bars can only drive retrograde motion in the vicinity of the bar (see their figure 5), i.e., within $\sim$5 kpc from the Galactic center of the Milky Way \citep{BlandHawthorn2016}.

The aforementioned facts indicate that BD+20 2457 is likely from the protodisk, but we do not completely rule out the possibility of it being an accreted star, due to the lack of studies of the chemical evolution of GE. Either way, this detailed chemodynamical analysis of BD+20 2457 origin has important implications to its mass and age as we will discuss in the following section.

\section{Mass and age of BD+20 2457}\label{sec:arr}

The BD+20 2457 system was discovered by \cite{Niedzielski2009}. According to the authors, the host star is a K2~II giant with 2.8 M$_\odot$ and the exoplanets BD+20 2457 b and BD+20 2457 c have 12.5 M$_{J}$ and 21.4 M$_{J}$, respectively. The host star has an apparent $G$-band magnitude $G = 9.234 \pm 0.003$ mag, $\varpi = 0.6787 \pm 0.02$\,mas, ($\mu_{\alpha^*}$,$\mu_{\delta}$) = ($-35.952\pm0.019$ ;$-29.96\pm0.018$) \,mas\,$\rm yr^{-1}$, and $\rm RV = 145.77 \pm0.19$\kms \citep{gaiaedr3}. Its atmospheric parameters are $T_{\rm eff} = 4258 \pm13$ K, $\log g = 1.64 \pm0.07$ (cgs), and $\rm[Fe/H]= -0.79 \pm 0.02$ dex \citep{Maldonado2013}. However, there is no consensus on the host star's mass in the literature; it is reported to vary from 0.8 to 10.83 $M_\odot$\citep{Zielinski2012,Stassun2017}. These differences in the stellar mass have important implications to the planetary masses reported originally by \cite{Niedzielski2009}.\par
We used \texttt{StarHorse} \citep{Queiroz2018} to estimate an isochrone mass and age for this star using the spectroscopic data of \citet{Maldonado2013}, the GEDR3 parallax, and multiwavelength photometric measurements from {\it Gaia}, 2MASS \citep{2MASS}, and AllWISE \citep{Wright2010} as input. When correcting for the parallax zero-point offset for this star (see Apeendix \ref{app:qcc}), we find that ages anywhere between 2 and 11 Gyr (masses between 0.9 and 1.5 $M_{\odot}$) are possible. The exact choices of the priors, the calibrations of parallax, surface gravity, and effective temperature, and their associated uncertainties, as well as considerable stellar model uncertainties in the upper red giant branch make it almost impossible to establish a reliable and precise age for BD+20 2457 without asteroseismic measurements. Nonetheless, we are still able to constrain its age and mass based on its chemistry and dynamics.\par
Firstly, we tested the hypothesis of BD+20 2457 as an old star through an estimation of the kinematic age of a population likely coeval with it. We use the same monoabundance population selected from the \textit{APOGEE sample} (see Section \ref{sec:BDacc}) and confirm that their velocitiy dispersions are commensurate with an old population, i.e., age $>$ 8 Gyr, according to the age--velocity dispersion relations of the monoabundance populations in \citealt{Mackereth2019}. This confirms that stars with chemical abundances such as BD+20 2457 are very unlikely to be young, i.e., age $<2$ Gyr.\par
Second, as discussed in Section \ref{sec:BDacc}, stars with halo-like kinematics and thick-disk abundances are most likely part of the heated protodisk. Furthermore, according to the assumption that the Galaxy's last significant merger was GE \citep{grand2020}, the protodisk stars were heated during this event or before, e.g. during a clumpy phase of disk. Therefore, we can use the epoch of the GE event to constrain BD+20 2457's age, and thus its mass. Several authors, using observations and/or simulations, converge on an estimated epoch for merger event between 9 and 11 Gyr ago (e.g., \citealt{belokurov2018, helmi2018,Bonaca2020}). As the dynamical effects of the merger on the protodisk can last for $\sim$1 Gyr \citep{bignone2019, grand2020,Bonaca2020}, we suggest that BD+20 2457 must be older than 8 Gyr, commensurate with direct age estimation of GE stars from \citep{gallart2019, montaban2020}.\par
Finally, based on the chemodynamical arguments and with the host star's metallicity and [$\alpha$/Fe] \citep{Maldonado2013}, we obtained a synthetic isochrone from the \textit{Dartmouth Stellar Evolution Database} \citep{dotter2008} and constrained the host star's mass to an upper limit of 1.00 $M_\odot$\footnote{The isochrone does not take into account the mass lost by the star over its lifetime.}. This results on planetary masses of $m\sin i \sim$ 13 and 7\,M$_{J}$ for BD+20 2457 b and BD+20 2457 c, respectively. These values are $\sim$50\% smaller compared to \cite{Niedzielski2009}.

\section{Conclusions and Final Remarks}
\label{sec:conc}

In this Letter, we investigated the origin, \textit{in situ} or accreted, of known planet-host stars. We combined the NASA Exoplanet Archive with GEDR3 and large ground-based spectroscopic surveys (namely, APOGEE DR16, GALAH DR3, and LAMOST DR5) and calculated the host stars' kinematic and dynamical parameters. Among the large sample of the thin-disk stars (97.2\% of the sample), we were able to identify 42 (2.7\% of the sample) thick-disk planet-host stars based on our kinematic criteria (see Section \ref{sec:ana}) and, most interestingly, a star with halo-like kinematics and dynamics: BD+20 2457.\par

Although the orbit of BD+20 2457 is very similar to what is commonly associated with GE stars, its chemical abundances are typical to what is associated with the thick-disk of the MW (see Section \ref{sec:BDacc}). Based on this finding, we suggest that BD+20 2457 is most likely a star born in the MW's protodisk. But, we cannot rule out the possibility of this star being formed in the GE as its chemical evolution is yet to be well understood. \par

The association of BD+20 2457 with the Galactic protodisk, or even with GE, has an important implication to its planetary system: it can constrain the host star's age and, therefore, its mass, which are not well defined in the literature (see Section \ref{sec:arr}).Moreover, the current mass estimation indicates that the BD+20 2457 system is dynamically unstable on a short time scale \citep{horner2014}. Then, we took into account the epoch of the GE merging event and were able to constrain the minimum age and the mass' upper limit of BD+20 2457: 8 Gyr and 1.00$M_{\odot}$, respectively. This decreases the estimated masses of its planets BD+20 2457 b and BD+20 2457 c in about 50\%.

Our prediction can be tested once asteroseismological data for BD+20 2457 become available, enabling more precise estimates for its mass and age. We note that TESS is due to observe this star in 2021 November\footnote{According to \url{https://heasarc.gsfc.nasa.gov/cgi-bin/tess/webtess/wtv.py}}. 

While this Letter was being written, \cite{Chen2021} presented a study that also classified planet-host stars into different Galactic components according to their kinematics. Their selection criteria removed BD+20 2457 due to its relative parallax error being greater than 10\% in \textit{Gaia} DR2. In GEDR3, whose data was employed in this work, the relative parallax error is $<$ 3\%. They also classified Kapteyn’s star (HD 33793) as a halo member, but this object was outside the cross-match range between NASA Exoplanet Archive and GEDR3 (see Appendix \ref{app:qcc}). Despite that, we investigated the Kapteyn’s star and it also has a heated disk-like orbit -- with a much lower vertical excursion from the Galactic plane ($Z_{\max} = 0.82$ kpc) compared to BD+20 2457 -- corroborate with its chemical properties \citep{woolf2004} which is more likely to belong to the heated disk.

This Letter provides a clear demonstration that, thanks to the \textit{Gaia} mission and large-scale spectroscopic surveys, extragalactic exoplanets can already be found inhabiting the Galactic stellar halo. Despite the origin of BD+20 2457 remaining unclear, 
our analysis inaugurates the exciting possibility of searching for exoplanet-host stars with chemodynamical signatures similar to that of accreted populations. Stars in the nearby halo have the enormous advantage of being much closer, hence significantly brighter, than any surviving satellite galaxy, thus allowing for immediate asteroseismic and/or high-resolution spectroscopic follow-up. With this approach, studies of the phenomena of planet formation in different environments will certainly be facilitated. We note that there are many millions of TESS targets still to be observed, thousands of which overlap with the kinematic footprint of GE \citep{carrillo2020}. Our group is currently working on the construction of a catalogue of likely-accreted TESS targets to be made publicly available to the astronomical community. In this scenario, it is reasonable to conjecture that the first extragalactic exoplanet, even if BD+20 2457 is not one such system, will be discovered in the Milky Way itself in the upcoming years.

\acknowledgments
We thank the anonymous referee for useful comments that helped to improve this work.
We also thank Thaise Rodrigues for helpful discussions about Bayesian methods to estimate stellar ages; Rodrigo Boufleur for  discussions about exoplanets masses; and Daniel Holdsworth for discussions about TESS mission. 
H.D.P. thanks FAPESP proc. 2018/21250-9, G.L. acknowledges CAPES (PROEX; Proc. 88887.481172/2020-00). S.R. would like to acknowledge support from FAPESP (Proc. 2015/50374-0 and 2014/18100-4), CAPES, and CNPq. F.A. acknowledges financial support from MICINN (Spain) through the Juan de la Cierva-Incorporci\'{o}n program under contract IJC2019-04862-I. L.B. acknowledges CNPq/PIBIC (Proc.136092/2020-9).

This research has made use of the NASA Exoplanet Archive, which is operated by the California Institute of Technology, under contract with the National Aeronautics and Space Administration under the Exoplanet Exploration Program.

This research has been conducted despite the ongoing dismantling of the Brazilian scientific system. 

\smallskip

\bibliographystyle{aasjournal}


\bibliography{bibliography.bib}

\appendix
\section{Quality control cuts}
\label{app:qcc}
We impose astrometric and spectroscopic quality cuts to obtain more precise and reliable orbital parameters. Below, we describe the adopted quality control cuts for each catalog data. 

\begin{enumerate}
\item We selected only {\it Gaia} EDR3 stars with high-quality parallaxes  ($\varpi/\sigma_{\varpi}>5$) and with good astrometric solutions  (\texttt{RUWE} $<1.4$; \citealt{Lindegren2020a}). We computed the distances as $d = 1/(\varpi)$ considering an offset of $-0.017$ mas (\citealt{Lindegren2020b}).

\item  We discarded APOGEE DR16 sources with bad spectroscopic flags {\tt ASPCAPFLAG != 0} and {\tt STARFLAGS = SUSPECT\_RV\_COMBINATION} to select stars with good estimates of radial velocities.

\item  We selected only GALAH DR3 stars with the flag parameter {\tt flag\_sp = 0} to ensure the quality of spectra and data.

\item  We obtained the LAMOST data from the added-value catalog of \cite{LAMOSTDDPayne}.
We selected stars with $r$-band signal-to-noise ratio (S/N) larger than 5, $g$-band S/N larger than 10, $i$-band S/N larger than $>$ 10 and quality flag {\tt qflag\_chi2 = good} and {\tt qflag\_singlestar = YES}.

\item  The \textit{APOGEE sample} has the same flags of item 2 and we applied additional cuts ($4000 < T_{\rm eff} < 6000$, $1 < \log g < 3$ and S/N larger than 70) in order to select only giant stars with high-quality elemental abundances.

\end{enumerate}
In order to build the samples, we performed an 3 arcsec radius cross-match between GEDR3 and the catalogs using TOPCAT (\citealt{TOPCAT2005}).

\section{Kinematical Selection Criteria}
\label{app:formalism}
We have adapted the formalism by \citet{bensby2003} for the discrimination of stars according to their Galactic component to the cylindrical coordinate system $(v_R, v_{\phi}, v_z)$. For nearby stars, this adaptation is straightforward since the $(v_R, v_{\phi}, v_z)$ points to the same directions of the $(U, V, W)$ vector in the Solar neighborhood.

As \citet{bensby2003}, we use a velocity ellipsoid for the distribution of the joint velocity probability distribution function for each Galactic component:
\begin{equation}
    f(v_R, v_{\phi}, v_z) = k \exp \left(-\cfrac{v_R^2}{2\sigma^2_{v_R}} -              \cfrac{(v_\phi-v_c-V_{\rm asym})^2}{2\sigma^2_{v_\phi}} -\cfrac{v_z^2}{2\sigma^2_{v_z}}
    \right)
    \label{bensby}
\end{equation}
where $k = \left([2\pi]^{3/2}\sigma_{v_R}\sigma_{v_\phi}\sigma_{v_z}\right)^{-1}$ is a normalization constant and $v_c$ is the local circular velocity. The values to be used in Equation \ref{bensby} for each Galactic component are given in Table \ref{ellipsoidvalues}, taken from \citet{Kordopatis2013} and \citet{amarante2020b}. This equation allows us to calculate the thick-disk-to-thin-disk (TD/D) and the thick-disk-to-halo (TD/H) membership ratios for each star using Eq. (3) from \citet{bensby2003}. We use these membership ratios to select thick disk and halo candidate stars from our sample of planet-host stars.

\begin{table*}[ht!]
\centering
\label{ellipsoidvalues}
\caption{Parameters for the Velocity Ellipsoid for Each Galactic Component.}
\begin{tabular}{lccccc}
\hline
\hline
       &  $X$   &  $\sigma_{v_R}$ & $\sigma_{v_\phi}$ & $\sigma_{v_z}$ & $V_{\rm asym}$ \\
Thin disk (D) & 0.9301 & 30 & 20 & 18 & $-$28 \\
Thick disk (TD) & 0.0652 & 61 & 45 & 44 & $-$63 \\
Halo (H) & 0.0047 & 160 & 119 & 110 & $-$228 \\
\hline
\end{tabular}
\end{table*}

\section{Catalog of Thick-Disk and Halo Candidates}
\label{app:exoplanets}

Table 2 provides a list of planets-host stars classified as thick-disk or halo stars. The columns {\it Gaia} EDR3 ID, RA, DEC, and Distance (see Appendix \ref{app:qcc}) are obtained from GEDR3. The columns RV, [Fe/H], $T_{\rm eff}$ and log $g$ are obtained from the surveys indicated by the symbol over the RV value. The Classification (Class.) indicates the Galactic component and the kinematic method used to classify it. The other columns are the derived orbital parameters, obtained as explained in Section \ref{sec:data}.

\begin{figure}
\includegraphics[width=\columnwidth]{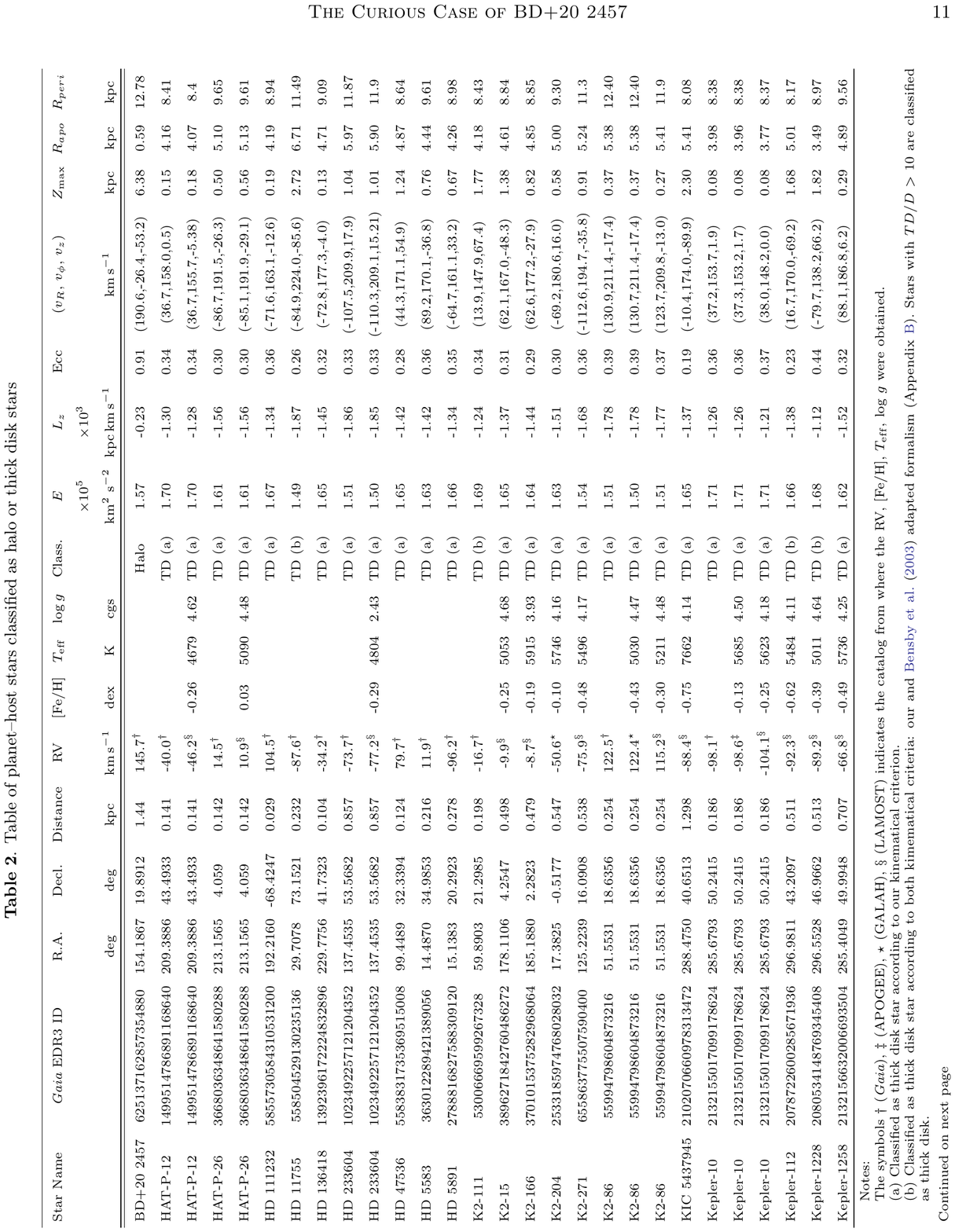}
\end{figure}
\begin{figure}
\includegraphics[width=\columnwidth]{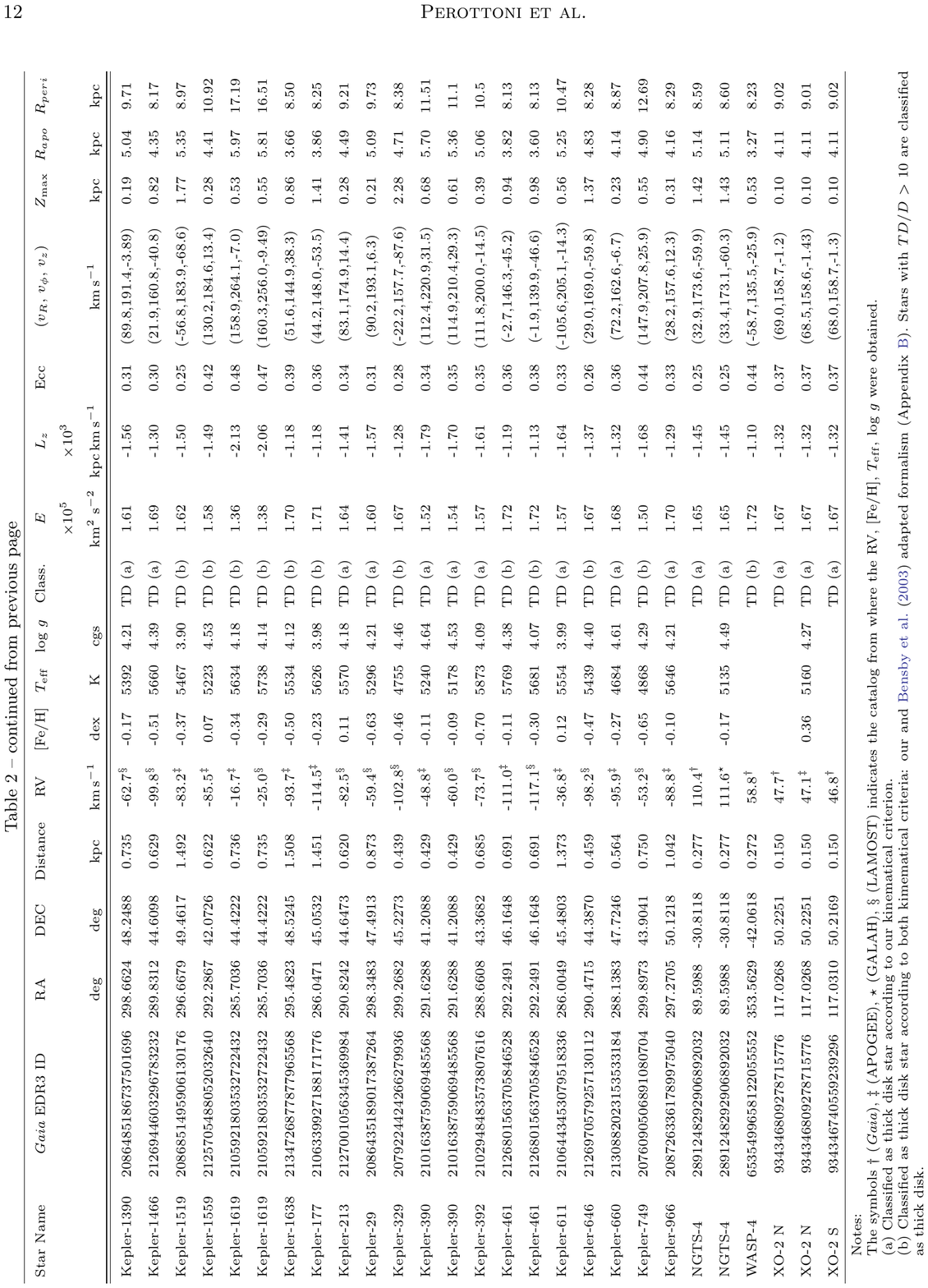}
\end{figure}


\setcounter{table}{0}

\end{document}